\documentstyle[aps,preprint]{revtex}
\input amssym.def
\input amssym.tex
\everymath{\rm}
\everydisplay{\rm}
\begin{document}
\title
{Double Exchange Alone Does Not Explain the Resistivity of $La_{1-x}Sr_xMnO_3$}
\author{A. J. Millis \\
P. B. Littlewood \\ B. I. Shraiman}
\address{AT\&T Bell Laboratories \\
Murray Hill, NJ 07974}
\maketitle
\renewcommand{\baselinestretch}{2}
\begin{abstract}

The $La_{1-x} Sr_x MnO_3$ system with
$0.2 \lesssim x \lesssim 0.4$ has
traditionally been modelled with a ``double
exchange'' Hamiltonian, in which it is assumed that
the only relevant physics is the tendency of carrier hopping
to line up neighboring spins.
We present a solution of the double exchange model, show it
is incompatible with many aspects of the resistivity data,
and propose that a strong electron-phonon interaction arising from
a Jahn-Teller splitting of the outer Mn d-level
plays a crucial role.
\end{abstract}
\newpage

The $La_{1-x}A_xMnO_3$ system (A represents
a divalent alkali element such as Sr
or Ca) has attracted much recent attention because
of the very large magnetoresistance exhibited for
$0.2 \lesssim x \lesssim 0.4$ \cite{GMR}.
Treatments\cite{Searle,Kubo,Furukawa} of the
physics of this system have focussed primarily on the
phenomenon of ``double exchange'' \cite{dexch}.
In this communication we show that a straightforward and
reasonably complete solution of the double exchange model
is possible.
{}From this solution we calculate the frequency and
temperature dependence of the conductivity
and show that it disagrees in several important respects
with the experimental data. The discrepancy must be resolved by including
additional physics, which we
suggest involves polaron effects due to
a very strong
electron-phonon coupling coming from a Jahn-Teller splitting
of the $Mn^{3+}$ ion. We
mention several experiments which have apparently not yet
been performed and which would verify or falsify our suggestion.

In $La_{1-x}A_xMnO_3$ the electronically active
orbitals are the $Mn$ $d_{x^2-y^2}$ and
$d_{3z^2-r^2}$ orbitals \cite{orbitals}.
The mean number of d-electrons per $Mn$ is 4-x,
the Hunds rule coupling is believed to be very strong relative to the d-d
hopping and the spin-orbit
coupling \cite{review} so the
spins of all of the d-electrons on a given site must be parallel.
Three of the d electrons go into tightly bound core-like
$d_{xy}, d_{xz}, d_{yz}$ orbitals forming a
core spin $S_i^c$ of magnitude 3/2,
to which the outer shell electron (which may hop from site
to site) is aligned by the Hund's rule coupling.
The Hamiltonian containing this physics
is
\begin{equation}
H_{d-ex} =-
\sum_{\langle ij \rangle a,b,\alpha}
t_{ij}^{ab}
d_{ia\alpha}^{\dagger} d_{jb\alpha}
- J_H
\sum_{i\alpha \beta}
\vec{S}_i^c \cdot
d_{ia\alpha}^{\dagger} \vec{\sigma}_{\alpha \beta}
d_{ia\beta}
\end{equation}

Here $d_{ia\alpha}^{\dagger}$ creates an electron in an outer-shell orbital
state $a = x^2 -y^2$ or $3z^2 -r^2$ and
spin $\alpha$, $J_H$ is the Hunds rule coupling connecting the
core spin to the outer shell electrons, and the interesting
limit, which we shall take, is $J_H \rightarrow \infty$.
To study Eq. (1) it is convenient to parametrize $S_i^c$ by
polar angles $\theta_i$, $\phi_i$ and to rotate the electrons
so that the spin quantization axis at, site $i$ is parallel
to $S_i^c$ on site $i$ and then project
on to the component parallel to $\vec{S}_i^c$.
The matrix ${\bf R}_i$ which accomplishes this is
${\bf R}_i = cos \theta_i / 2$
${\bf 1} + i sin ( \theta_i /2) sin \phi_i  \mbox{\boldmath $\sigma$}^x + i sin
( \theta_i /2) cos \phi_i  \mbox{\boldmath $\sigma$ }^y$.
The electrons may be integrated out, and the partition function $Z$
written as
\begin{equation}
Z = \int {\cal D} cos \theta {\cal D} \phi_i \; exp \cal A
\end{equation}
with the action ${\cal A}$ given by
\begin{equation}
\begin{array}{lr}
{\cal A} =  & Tr \; ln \:
[ \partial_{\tau} - \mu +
R_i^+ \partial_{\tau}
R_i - t_{ij}^{ab}  (R_i^{\dagger}
R_{j} + R_J^{\dagger}R_i )]_{11} \\
  &+ i S_c \int_0^B d \tau \sum_i \dot{\phi}_i
(1-cos \theta_i )
\end{array}
\end{equation}
Here the first term comes from integrating out
the electrons and the second is the Berry phase
term for the core spins.
The subscript 11 on the argument of the logarithm
comes from the requirement that in the $J_H \rightarrow \infty$ limit
the outer shell electron must be parallel to $\vec{S}_c$.
At low $T$ we may expand
about the ordered ferromagnetic state
$\theta_i = 0$.
The effective action becomes
${\cal A} = {\cal A}_F + {\cal A}_{sw}$. ${\cal A}_F$ is the free energy
for free fermions moving in the band structure defined by $t_{ij}^{ab}$.
${\cal A}_{sw}$ may be written in terms of the
magnetization variables $M_x$ and $M_y$ describing
deviations from the ordered state (with magnetization taken parallel to z) as
\begin{equation}
\begin{array}{rl}
{\cal A}_{sw} & = \frac{1}{2}
\int_0^B d \tau
\int \frac{d^3k}{(2\pi)^3}
i \left ( S_c + \frac{1-x}{2} \right )
\vec{M}_k \times \partial_{\tau}
\vec{M}_{-k} \nonumber \\
  & - 2Ka^2k^2 \vec{M}_k \cdot \vec{M}_{-k} + {\cal O} (M^4)
\end{array}
\end{equation}
Here $K= \sum_{ab} t_{i, i+ \hat{x}}^{ab}
\langle c_{ia}^\dagger c_{i+ \hat{x}b} \rangle$ is the electron stress-
energy tensor and is related to the integral of the optical conductivity
as described below.
$a$ is the lattice constant.  Here $c_i^{\dagger}$ creates a spin polarized
electron on site i.
Equation (4) is the action for a quantum ferromagnet
with spin $S^*=S_c+ \frac{1-x}{2}$
and stiffness $K$.
This action implies that the magnon dispersion is
\begin{equation}
\omega_{mag} = (K/S^* ) (ka)^2
\end{equation}
A very similar result for the magnon dispersion was obtained by Kubo and Ohata
\cite{Kubo} using different methods.

The quantity $K$ is very important because it is the only
energy scale in the theory.    We may estimate $K \sim 2tn$, where $n$ is the
electron density and $t$ the hopping energy.  A recent band theory
calculation found a bandwidth of $2eV$ implying $t \sim 0.25eV$ and
$K \sim 0.1 eV$ for $n = 0.3$.
\cite{Mattheiss}.From $K$ we may, e.g. estimate
the ferromagnetic transition temperature  as follows:
the cubic lattice Heisenberg model with exchange
constant $J$ has a magnon dispersion
$\omega = 2 JS (ka)^2$ \cite{Kittel}.  The known relation \cite{Domb}
between $J$ and
$T_c$  then implies a $T_c \approx
2.9 K (S^*+1)/S^* \approx .3 eV$, more than an order of magnitude
higher than the observed value.  This discrepancy, we believe, is evidence
that additional physics, not included in the double exchange model,
is important for $La_{1-x}Sr_xMnO_3$.

A direct measurement of the magnon spectrum would determine $K$.
In the absence of this measurement, one may
estimate $K$ from the optical conductivity $\sigma ( \omega )$.
In a one-band model with only nearest neighbor hopping
$\int_0^{\infty} d \omega \sigma ( \omega ) =
\pi e^2 K/a$\cite{optics}.
To extract $K$ from conductivity data on a real material one must
remove the interband contributions to $\sigma$.   Ambiguities arise because
there is often
no clear demarcation between interband and intraband
contributions.
The optical conductivity of
$La_{1-x}Sr_xMnO_3$ has been measured for
$x=0.175$ and $x=0.3$\cite{optics data}.
Roughly,
$\sigma (\omega) \approx (500 \mu \Omega -cm)^{-1} \approx 0.3$
eV, independent of $\omega$ and it seems reasonable to assume that for
$\omega <1 eV$ the conductivity is dominated by the conduction band.
Using $a = 4 \AA$, one finds
$K \approx 0.03 \; eV \approx 400K$, much less than the band structure
estimate.
Note that even this value of K implies a magnetic
transition temperature.
 much higher than the observed
$T_c \approx 200K$ for $x=0.175$.

We now turn to the properties of the model in the regime
$T \sim T_c$.
Because we have already shown that the low $T$
properties are those of a quantum model with
the relatively large spin value
$S \approx 2$,
it seems reasonable to suppose that near $T_c$
we may consider classical spins and so neglect
the imaginary time dependence of the angular variables
in Eq. (3). The problem then becomes that of electrons of
electrons moving on a lattice with hopping amplitude
$\bar{t}_{ij}^{ab} = t_{ij}^{ab} [ cos (\theta_i/2) cos (\theta_j/2 )$
$+ cos (\phi_i-\phi_{j}) sin (\theta_i/2) sin (\theta_{j} / 2)]$.
We further assume that contributions to the partition function
in which fermions move on closed loops in real space may
be neglected. We may then
rotate the $\phi_i$ independently and therefore
replace this by the familiar double exchange form
$\bar{t}_{ij}^{ab} =t_{ij}^{ab} \sqrt\frac{S^2+\vec{S}_i \cdot
\vec{S}_{j}}{2S^2}$\cite{dexch}.
We may then replace Eq. (1) by
\begin{equation}
H_{eff} =- \sum_{ij ab}
\frac{t_{ij}^{ab}}{\sqrt2}
\sqrt{1+ \frac{\vec{S}_i \cdot \vec{S}_{j}}{S^2}}
(c_{ia}^+ c_{j b} + h.c.)
\end{equation}
where the $\vec{S}_i$ are now understood to be
classical spins.
The free energy function describing the spin
distribution is to be obtained by integrating out the
conduction electrons.
For any fixed distribution of spins the problem
is one of conduction electrons moving in a lattice with random
hopping.
To a good approximation, the free energy of the conduction
electrons depends only on the average hopping\cite{Ziman} so the spin
energy $E( \{ S_i \} )$ is given by
\begin{equation}
E ( \{ S_i \} ) =-
T \sum_k ln \left [ 1+ e^{\beta ( \epsilon_k - \mu )} \right ]
\end{equation}
with
\begin{equation}
\bar{\epsilon}_k =- 2 \bar{t} (cos k_x a+ cosk_ya+cosk_za)
\end{equation}
and
\begin{equation}
\bar{t} = \langle t_{ij}^{ab}
\sqrt{1+\frac{S_i \cdot S_{j}}{S^2}}\rangle
\end{equation}
In particular, if the temperature is less than the Fermi
temperature of the electrons,
\begin{equation}
E ( \{ S_i \} ) =-
\sum_{\langle i j \rangle}
\frac{t_{ij}^{ab}}{\sqrt2}
\sqrt{1+\frac{S_i \cdot S_{j}}{S^2}}
\langle c_{ia}^+ c_{j b} \rangle
\end{equation}

In other words, the spin energy involves nearest neighbor coupling with
scale again set by the electron kinetic energy.
In the nearest neighbor Heisenberg model at
$T_c , \langle \vec{S}_i \cdot \vec{S}_{j} \rangle / S^2 \approx 1/3$
\cite{Domb}, so an expansion in
$S_i \cdot S/S^2$
is reasonable, and we may conclude that the spin energy is
given by the nearest neighbor Heisenberg model, with
$J = K / 2 \sqrt2$.  Thus thermal effects do not significantly
change our estimates of the energy scales, which are much too
large to explain the observed transition temperature.

We now turn to the resistivity of the model near $T_c$.
Before presenting the details of the calculations, we make some general
comments.
The electron-spin-fluctuation interaction in Eq. (6)
leads to an electron self-energy with real and imaginary parts.
The real part leads to a contribution to the electron
velocity which increases as the temperature decreases and
expresses the physics that as
$\langle \vec{S}_i \cdot \vec{S}_{j} \rangle$
increases, so does the electron hopping.
The imaginary part leads to scattering, due physically
to fluctuations in $\vec{S}_i \cdot \vec{S}_{j}$.
If the spins are treated classically, this scattering is
mathematically identical to conventional impurity
scattering, and leads to a resistivity proportional to
$(p_F \ell )^{-1}$ where $p_F$ is the electron Fermi
wavevector and $\ell$ is the mean free path.
The mean free path in this static spin approximation is a purely geometric
property
determined by the amplitude and spatial correlations
of the fluctuations in $\vec{S}_i \cdot \vec{S}_{j}$
and in particular is independent of the electron velocity.
Therefore to compute the resistivity it suffices to
calculate the scattering of the electrons off the spin
fluctuations, neglecting the possibly large velocity renormalization.
Further, we shall see that the scattering is sufficiently
weak that the Born approximation suffices.
Finally, we note that because the scattering is static
the optical conductivity must have essentially
the Drude form.

There is no rigorous expression for the dc resistivity.
To obtain a reasonable approximate expression we
use the convenient and at least qualitatively
accurate ``memory function'' method\cite{memory}
in which one defines the memory function $M( \omega ,T)$ via
\begin{equation}
M ( \omega , T) =
\int_0^{\infty} dt \;
e^{i \omega t}
\langle [ H , j ]_t , [H, j ]_0 \rangle
\end{equation}
where the current operator $j$ is
\begin{equation}
j = i \sum_{i j ab}
t_{i j}^{ab}
(c_{ia}^+ c_{j b} - c_{j b}^+ c_{ia})(1+frac{S_i \cdot S_j} {S^2})
\end{equation}
and $H$ is given by Eq. (6).
The Heisenberg representation is assumed
and the subscript on the commutator denotes the time
argument of the operators.
The memory function rigorously determines the leading term in a
high-frequency expansion of the conductivity.
By assuming that this leading term is the first
term of an expansion of $\sigma ( \omega , T ) =Ke^2 /a$
$\left [ i \omega+ \frac{M(\omega , T)}{K} \right ]$
one finds \cite{memory} that the temperature dependent resistivity
$\rho(T) = e^2 M ( \omega =0,T)$.
We have evaluated $M(\omega =0 ,T)$ from Eqs. 6 and 11 to leading order in
$1/S$ and $k_Fa$.
We find
\begin{equation}
\rho(T) = e^2
\sum_{R, \delta_1, \delta_2}
\langle \vec{S}(0) \cdot
\vec{S} (- \vec{\delta}_1 )
\vec{S}(R) \cdot
\vec{S}(R+ \delta_2 ) \rangle /S^4
B(R)
\end{equation}
Here $\vec{R}$ labels sites on the cubic lattice, and $\delta_1$ and $\delta_2$
are any of the vectors
$\hat{x} , \hat{y} , \hat{z}$ connecting a site to one of the nearest
neighbors.
$B$ is proportional to the electron current-current
correlation function weighted by a factor accounting for the
ineffectiveness of small $q$ scattering in degrading the current.
In the free electron approximation in which the fermions have
a $k^2$ dispersion and $t^a=t^b=t$,
\begin{equation}
B(R) =
\frac{9}{32(p_Fa)^4}
\left [ \frac{sin^2p_F ( \vec{R}+ \hat{x})}{(p_F | \vec{R}+ \hat{x} |)^2}+
\frac{sin^2p_F | \vec{R}- \hat{x}|}{(p_F|R - \hat{x}|)^2}
- \frac{2sin^2p_F R}{(p_F R)^2} \right ]
\end{equation}
Here $p_F$ is the Fermi wavevector.

It is interesting to compare Eqs. 11 and 12 to
the expression for $\rho (T)$ given by Langer
and Fisher\cite{langer fisher}
who considered the general question of resistive anomalies
at magnetic critical points.
They began from a model in which the carrier-spin coupling
was $H_{c-=s} \sum_i \vec{S}_i \cdot \vec{\sigma}_{ci}$ where $S_i$
is a local moment and $\sigma_c$ the carrier spin density at site $i$,
and obtained a formula rather similar to Eqs. 11, 12 except that instead of
the four spin correlator they obtained simply
$\langle \vec{S} (0) \cdot \vec{S} (R) \rangle$ because in
their model, local fluctuations of $S_i$
scatter the electrons while in the model defined by Eq. 6
local fluctuations of $ S_i \cdot S_j $
scatter the electrons.
Langer
and Fisher found that there are two sorts of  resistive anomalies at
a ferromagnetic transition:
for $T > T_c$,
$d \rho / dT \sim (T-T_c)^{- \alpha}$
where $\alpha$ is the specific heat exponent (which is
believed to be slightly negative for the three
dimensional Heisenberg model) while for
$T < T_c$ there is an additional contribution to the
resistivity proportional to the square of the magnetization.
These general conclusions apply also to the present model.

One important consequence of the difference in models is that
the peak at $T>T_c$ found for small $k_F$ by Langer
and Fisher is absent in the present model.
The point is that the function $B$ has range
$k_F^{-1}$.
In the range where this is greater than the magnetic
correlation length, the two spin correlator in the Langer-Fisher expression for
$\rho$
diverges in the same way as
the uniform susceptibility, i.e. as $\xi^4$
In the present case,
the quantity in square brackets is the long-wavelength fluctuations
in $S_i \cdot S_{j}$, which has the
weaker divergence
$\xi^{4-d-2 \eta}$.
In the calculations we have performed for
$0.5 \leq k_F a \leq 1$ the divergence is not visible.

For $T <T_c$, or in a magnetic field the same formula
applies except that one must distinguish between longitudinal and transverse
fluctuations, and one must add terms in which two of the
spins in Eq. 11 are replaced by the uniform magnetization.
These terms lead to a contribution to $\rho$ proportional to $M^2$ which is
absent in the model of Langer and Fisher.
In particular, in the model of Langer and Fisher, the only effect
of a nonzero $M$ was to decrease the total amplitude
of the spin fluctuations, leading to a decrease in $\rho$
below $T_c$.
In the present model, this effect competes with the four additional
scattering terms of form
$M^2 \langle S(0) \cdot S (R) \rangle$ which lead to an
increase in the resistivity.

We have evaluated Eq (11) using Eq. 12 for $B$ and
calculating the spin correlator in the
spherical model.
Results are shown in Fig. 1 for $p_Fa =1$
and two magnetic fields:
$H = 0$ and $H = 0.1 T_c$.  Results for $p_Fa =0.5$ are very similar.
Note that the results are consistent with the predictions of
Langer and Fisher:
in the spherical model $\alpha =-1$ so $C$ and
$d \rho / dT$ have derivative discontinuities at
$T_c$ and below $T_c$ an additional term, proportional to
$M^2$, is operative.
However, in the spherical model approximation used here the sign
of this term is positive:
the resistivity {\it increases} below $T_c$ or in a field.

The resistivity implied by Eq. 6 has been previously
calculated by Kubo and Ohata \cite{Kubo}, Searle and Wang \cite{Searle},
and Furukawa \cite{Furukawa}.
Searle and Wang and Furukawa used mean field approximations in which
all spin correlations are neglected, i.e.
$\langle S_i \cdot S_j \rangle - \langle S_i \rangle$
$\langle S_j \rangle = 0$.
It is evident from the previous discussion that these
correlations are essential.
Furukawa used an "infinite dimensional" approximation in which he found that
for
$T > T_c$ the core spins fluctuated very rapidly (i.e. on the scale set by t)
and led to an enormous imaginary part ($\approx J_H$) to the electron self
energy.  It is difficult to reconcile these results with those presented
here.  In our work $J_H$ drops out of the problem and the $S_i$ are seen to
be well described near $T_c$ by a classical Heisenberg model which entails
fluctuation rates of order T or less.
Kubo and Ohata obtained via a different method an expression
very similar to our Eq. 18, but evaluated the spin correlation
function using an approximation which neglected the fluctuations in
$\langle S_i \cdot S_j \rangle$ which are responsible for the up-turn
we find in $\rho$ near $T_c$.

Some representative experimental data from ref \cite{resistivity}
are shown in the inset to Fig. 1.  Similar experimental results have been
obtained
by many authors \cite{GMR,Searle,kusters,schiffer}.
Although the qualitative temperature dependence calculated
for $T > T_c$ is consistent with the data several important
discrepancies are evident:
the calculated resistivity has the wrong magnitude
(by several orders of magnitude), a far too weak doping
dependence, and an incorrect behavior
for $T <T_c$ or in a field.
Some of these discrepancies may be due to
the inadequacy of the spherical model, but the
magnitude and doping dependence cannot easily be explained
away.
The results may be traced to the fact, evident already in Eqs. 9
and 10, that in the double exchange model the magnetic fluctuations are
a weak perturbation on the electron hopping and imply a
$k_F \ell \gg 1$.

The discrepancy suggests that some other mechanism, not present in the
double exchange model, must act to substantially
reduce the electron hopping.
We suggest that  this mechanism is a polaron effect due to a very strong
electron-phonon coupling stemming from a Jahn-Teller splitting
of the $Mn^{3+}$ ion.  Other authors, most notably Kusters et. al.
\cite{kusters} have argued in favor of a magnetic polaron picture.
Our calculation shows that the standard double exchange Hamiltonian
does not contain magnetic polaron effects because the effective
carrier-spin interaction is
too weak to  lead to the formation of polarons.
On the other hand, the  Jahn-Teller coupling is very strong.  It causes the
cubic-tetragonal
transition observed at $T^* \approx 800K$ in $LaMnO_3$\cite{orbitals}, and
in fact, $T^*$ is a dramatic underestimate of the basic Jahn-Teller
energy.
Using the standard Jahn-Teller Hamiltonian and the
measured\cite{oxygen} oxygen displacements one finds
that this energy is $\sim 1 \: eV$, much greater than the measured electron
kinetic energy at $x = 0.175$.
It therefore seems likely that the Jahn-Teller energy
remains important even in the metallic regime
$0.2 < x < .45$.
In this picture the physics would involve a crossover between
a high $T$, polaron dominated disordered regime and a low
$T$ metallic ordered regime.
Understanding this crossover requires a theory of the
interplay of polaron and metallic physics which is beyond the scope of this
paper.
However, it is clear that if polaron
physics reduces the mean free path to less than a lattice constant,
the argument previously given that the
carrier resistivity is independent of the mass does not apply.
Mathematically, if the electron self-energy is very large,
than the hopping part of the electron Green function
is proportional to $t$, not $1/t$ and so the memory function,
Eq. (11) scales as $t^{(4)}$, not $t^{(0)}$.
Then the increase in velocity for $T < T_c$ or $H \neq 0$
will compete with the extra scattering terms and
may lead to drop in resistivity for $T <T_c$, as observed.

In conclusion, we have presented and compared to data a solution of the
``double exchange'' Hamiltonian widely believed to describe
the physics of $La_{1-x} Sr_x MnO_3$.
We noted the existence of a relation between an
optical property (the low frequency spectral weight) and a
magnetic property (the spin-wave stiffness).
Experimental measurements of the spin wave stiffness would
be a useful test of the model.
We showed that the calculated resistivity is much too small, and
has an incorrect field and temperature dependence, and we proposed that polaron
effects are responsible for the discrepancy.
\section*{Acknowledgements}
We thank Dr. D. E. Cox for drawing our attention to ref \cite{oxygen}
and Dr. P. Schiffer for many interesting discussions of the data and for
drawing
our attention to ref \cite{kusters}.

\newpage
\begin{figure}
\caption{
Resistivity calculated from double exchange model as described in the text.
The solid line is the resistivity in zero field; the dashed line is the
resistivity in a field of 0.15$T_c$.  The inset displays data from Tokura
et. al..
}
\end{figure}

\newpage
\begin{references}
\bibitem{GMR}
S. Jin, T. H. Tiefel, M. McCormack,
R. A. Fastnacht, R. Ramesh and L. H. Chen,
Science {\it 264}, 413 (1994).
\bibitem{Searle}
C. W. Searle and S. T. Wang,
Canadian Journal of Physics {\it48}, 2023 (1970).
\bibitem{Kubo}
K. Kubo and N. Ohata,
J. Phys. Soc. Jpn. {\it 33}, 21 (1972).
\bibitem{Furukawa}
N. Furukawa, J. Phys. Soc. Jpn, in press.
\bibitem{dexch}
C. Zener, Phys. Rev. {\it82}, 403 (1951), P. W. Anderson
and H. Hasegawa, Phys. Rev. {\it100}, 675 (1955),
P. G. deGennes, Phys. Rev. {\it 118}, 141 (1960).
\bibitem{orbitals}
J. Goodenough, Phys. Rev. {\it 100}, 564 (1955).
\bibitem{review}
C. Herring, {\it Magnetism}, vol 2,
J. Rado and H. Suhl, eds.
\bibitem{Mattheiss}
L.F. Mattheiss, private communication.
\bibitem{Kittel}
C. Kittel {\it Quantum Theory of Solids}
(Wiley, New York) 1963 Eq. 26 of Ch. 2.
\bibitem{Domb}
G. S. Rushbrooke, G. A. Baker, Jr. and P. J. Wood in {\it Phase Transitions and
Critical Phenomena} C. Domb and M. S. Green, eds., (Academic Press:  New York)
1974.  See especially eq 5.4.
\bibitem{optics}
A. J. Millis and S. N. Coppersmith,
Phys. Rev. B{\it 42}< 10807 (1990).
\bibitem{optics data}
T. Arima,  unpublished.
\bibitem{Ziman}
J. M. Ziman {\it Models of Disorder}.
\bibitem{memory}
W. Gotze and P. Wolfe, Phys. Rev. B{\it 6}, 1226 (1972).
\bibitem{langer fisher}
M. E. Fisher and J. S. Langer,
Phys. Rev. Lett. {\it 20}, 665 (1968).
\bibitem{resistivity}
Y. Tokura, A. Urushibara, Y. Moritomo, T. Arima, A. Asamitsu, G. Kido and N.
Furukawa, unpublished.
\bibitem{kusters}
R. M. Kusters, J. Singleton, D. A. Keen, R. McGreevy and W. Hayes, Physica
{\it B155} p. 362 (1989).
\bibitem{schiffer}
P. Schiffer, A. P. Ramirez, W. Bao and S-W. Cheong, unpublished.
\bibitem{oxygen}
J. B. A. A. Elemans, B. vanLaar,
K. R. vanderVeer and B. O. Loopstra,
J. Sol. St. Chem. {\it 3}, p.238 (1971).
\end{references}
\end{document}